# On relative input of viscous shear into the elasticity equation at near-front, near-inlet and the major part of a hydraulic fracture

**A. M. Linkov**

Rzeszow University of Technology, Poland

e-mail: linkoval@prz.edu.pl

*Abstract.* The input of the hydraulically induced shear traction into the elasticity equation is estimated for the entire fracture surface. It is established that, except for negligibly small vicinities of a pointed source and of the fluid front, the relative input of the viscous shear does not exceed $10^{-4}$ of the input of the conventionally accounted term. The estimation is true for Newtonian, as well as non-Newtonian thinning fluids. This implies that there is no need to account for viscous shear not only in the form of conventional near-front asymptotics and fracture conditions, but also in factors entering them.

*Key-words:* hydraulic fracture; viscous shear stress; elasticity equation

## 1. Introduction

The conventional formulation of a hydraulic fracture (HF) problem does not include shear stresses, induced by viscosity: they are already taken into account when deriving the Poiseuille type equation of a fluid motion (e.g. [1]). Still, in general, these stresses are present in the exact elasticity equation [2-5], which connects the fracture opening with the net-pressure. Normally, the shear tractions, which arise under usual fluid velocities, are much less than other terms entering the elasticity equation. For this reason, starting from the paper by Spence and Sharp [1, p. 291], these tractions are neglected when studying and modeling HF (e.g. [6-12]).

Recently Wrobel et al. [13] have noted that *near the fracture tip*, the limit of the shear traction to the net-pressure goes to infinity. This impelled authors to include the shear traction into the elasticity equation and into the fracture condition based on energy release rate. The starting statement, typed in bold in [13, p. 31], is: "The basic assumption of the classic HF theory is violated, at least near the crack front".

This *qualitative* conclusion has been made under the usual assumption that the lag between the fluid front and the fracture contour is small and may be neglected. Then points of the fluid front are singular and shear traction goes to infinity when approaching the front. The derivations of asymptotics, modified expressions for the energy release rate and fracture criterion, given in [13], refer utterly to the *near-front* zone, that is to a vicinity of these singular points. They are solely based on the starting statement cited. An example of the self-similar solution to a problem with exponentially growing pumping rate serves the authors to demonstrate that the input of the shear term in the near-front zone may reach 10%. However, a *quantitative* analysis [14,15] has shown that the input of shear traction into the elasticity equation in the near-front zone is negligible, while the example is odd.

There are other singular points of the flow, where the fluid velocity and consequently viscous traction at walls of a narrow channel, may go to infinity. Such are injection points simulated by using Dirac's delta function. This raises the question: if hydraulically induced shear tractions should be accounted for in the elasticity equation and fracture conditions when solving problems of hydraulic fracture propagation, at least in some special cases? Our objective is to give an unambiguous answer to the question.

## 2. General equations

Hydraulically induced shear tractions are not present explicitly in the equations describing a fluid flow in a narrow channel. Actually, these tractions are accounted for inexplicitly in the Poiseuille-type equation. The latter in terms of the shear at walls of a channel is written as [14]:

$$\tau = \frac{\mu'}{2}\left(\frac{v}{w}\right)^n, \quad (1)$$

where $\tau$ is the magnitude of shear traction, $v$ is the magnitude of the particle velocity, $w$ is the fracture opening, $\mu' = 2\left(2\frac{2n+1}{n}\right)^n$, $n$ and $M$ are, respectively, the fluid behavior and consistency indices. For a Newtonian fluid ($n = 1$), $M$ is the dynamic viscosity. Equation (1) is true *at any point* of a hydraulic fracture where the Poiseuille-type formula is applicable.

As mentioned in Introduction, despite the shear tractions $\tau$ are not present in fluid equations, they enter the exact boundary integral equations (BIE) for elastic rocks. To compare their input with that of terms, which are conventionally accounted for, one may use the BIE of the elasticity theory for 3D piece-wise homogeneous bodies with displacement and/or traction discontinuities at contacts of structural elements [3-5]. For the plane-strain problem involving a straight crack in a homogeneous plane, the input emmidiately follows from the classical Muskhelishvili's equation [2]. More involved derivation, given in Appendix for an arbitrary planar fracture, shows that the input is defined by the ratio similar to that for a straight crack:

$$R_\tau = \frac{2k_\tau \tau}{E'|\partial w/\partial x|}, \quad (2)$$

where $E' = E/(1 - v^2)$ is the plane-strain elasticity modulus, $E$ is the Young's modulus, $v$ is the Poisson's ratio; $\partial w/\partial x$ is the drivative of the opening in the direction of shear traction on the fracture surface; $k_\tau = \frac{1-2v}{1-v}$ is the factor depending merely on the Poisson's ratio. Its maximal value being 1, one may set $k_\tau = 1$ when estimating maximal values of the ratio $R_\tau$.

Substitution (1) into (2) yields the general equation for the relative input $R_\tau$ of the hydraulically induced shear tractions in the elsticity equation:

$$R_\tau = k_\tau \left(\frac{t_n v}{w}\right)^n \frac{1}{|\partial w/\partial x|} \quad (3)$$

Herein,

$$t_n = \left(\frac{\mu'}{E'}\right)^{1/n} \quad (4)$$

is the only constant with the time dimension, which is present in equations describing HF. Note that the typical order of this time ($10^{-11}$ s) is of key significance for evaluation the order of $R_\tau$.

**3. Input of viscous shear at near-front zone**

For the *near-front* zone, equation (1) may be written by using the front propagation speed $v_*$, rather than the particle velocity $v$ at an arbitrary point:

$$\tau = \frac{\mu'}{2}\left(\frac{v_*}{w}\right)^n \quad (5)$$

The derivation of the asymptotic equation (2) employs the non-trivial fact that, although the shear stress $\tau$ is singlar at the tip, the particle velocity $v$ is continous near the tip; it goes to the limit equal to the *finite non-zero* propagation speed $v_*$ due to the speed equation [16-19,11].

Using (5) in (2) gives the relative input $R_\tau$ for the near-front zone [14]:

$$R_\tau = k_\tau \left(\frac{t_n v_*}{w}\right)^n \frac{1}{dw/dr}, \quad (6)$$

where $r$ is the distance from the fluid front. In difference with (3), equation (6) contains the fracture propagation speed rather than the particle velocity.

The lag between the fluid front and the fracture contour being small, it is normally neglected when studying HF. Then the front and the contour coincide, and the points of the front become singular points, at which the shear traction goes to infinity. The asymptotic behavior of fields near a point at the front has been the subject of numerous studies (e.g. [20, 1, 21, 6, 22-24, 11]). In general, the asymptotics discussed in these papers are of a monomial form. They are summarized by the universal asymptotic umbrella (UAU) [19, 11]. The UAU expresses the opening through properties of the fluid and rock and the propagation speed. For actually arbitrary regime of the HF propagation, the UAU is represented by the almost monomial dependence:

$$w = A_w(v_*)r^\alpha, \tag{7}$$

where the exponent $\alpha$ and the form of the function $A_w(v_*)$ change very slow in a wide range of the propagation speed $v_*$. By using (7) in (6), the ratio becomes

$$R_\tau = k_\tau \frac{(t_n v_*)^n}{\alpha A_w^{n+1} r^{\alpha(n+1)-1}} \tag{8}$$

In the limiting cases of toughness, viscosity or leak-off dominated regimes, the dependence (7) reduces to classical asymptotic equations [20, 1, 6] for these regimes.

Clearly, the greatest impact of viscous shear occurs in the viscosity dominated regime. For this regime, $\alpha = 2/(n+2)$ and $A_w = A_\mu(t_n v_*)^{1-\alpha}$ with $A_\mu = [(1-\alpha)B(\alpha)]^{1/(n+2)}$, $B(\alpha) = \frac{\alpha}{4}\cot[\pi(1-\alpha)]$. Substitution $\alpha$ and $A_w$ into (8) yields:

$$R_\tau = k_\tau \frac{1}{\alpha A_\mu^{n+1}} \left(\frac{t_n v_*}{r}\right)^{n/(n+2)} \tag{9}$$

For typical values of fluid and rock parameters $n$, $\mu'$, $E'$ and for quite large value of the propagation speed $v_* = 0.1$ m/s, equation (9) implies that the relative input of the hydraulically induced shear traction in the elasticity equation near the front is negligible. It is less than 1% except for a zone of an atomic size. Surely, such a small zone is beyond applicability of continuum mechanics, physical significance, computational abilities of computers and practical applications of HF [14, 15].

The relative input rapidly decreases with growing distance $r$. From (9) it is easy to see that when the distance $r$ becomes close to the size of the zone under the UAU (7), the relative input $R_\tau$ decreases to the order $10^{-4}$ at most. Hence, there is no need to account for hydraulically induced shear tractions in asymptotic equations, following from the conventional analyses. This also implies that there is no need to modify the *form* of conventional fracture conditions (cf. [25, 26-28]) by including into them a special term, which accounts for these tractions. Summarizing, it is of no sense to account for the input of *near-tip* hydraulically induced shear tractions in the elasticity equation, in the equation for the energy release rate and in the form of fracture criterions.

*Comment 1*. In the example of the paper [13], the authors obtained numerical results, which drastically differ from the estimations above following from (9). Under the assumption of *exponentially growing pumping rate*, in the case of the viscosity dominated regime they had the ratio $R_\tau$ near the front of order 1. According to (9), such an input of the shear term into the elasticity equation may occur merely for extremely high values of the fracture propagation speed $v_*$. The detailed analysis given in [14, 15] shows that indeed the speed $v_*$, corresponding to the exponential influx, is tremendously high. Specifically, according to the solution given in [13], if at some instant, the pumping rate has an order typical for hydraulic fracturing, then at this instant, the fracture propagation speed $v_*$ is four orders greater, than values typical in practice. Even more extraordinary is, that during extremely short time interval (of order $10^{-8}$ s) after this instant, the fracture propagation speed, given by the self-similar solution, exceeds the speed of light. Besides, at the only instant, when the flux has the physically sensible value, the fracture length is enormously small

44($6 \cdot 10^{-7}$ m). This unrealisitc size is an additional evidence that the exponential solution is exotic, to say the least. Clearly, it cannot support any statement on the influence of the hydraulically induced shear stresses.

Out the asymptotic zone, the shear traction may influence merely unknown *factors*, which enter conventional asymptotics describing quantities at this zone. These factors are defined by the global solution, and naturally, they may depend on shear stresses out of the near-tip zone. Therefore, it is of interest to study the input of viscous shear into the elasticity equation beyond this zone.

**4. Input of viscous shear at near-inlet zone**

Commonly HF propagation is studied separately from detailed modeling of complicated flow through the perforation in the near borehole area. The injection is represented by a pointed source $q_0(t)\delta(x)$ with $q_0(t)$ being the pumping rate and $\delta(x)$ the Dirac's delta-function (e.g. [1, 7-13, 24, 29]). Then in 3D problems, the mass conservation law, applied to a flow of incompressible fluid in a narrow channel, yields [8, 29] that the particle velocity goes to infinity near the inlet as

$$v = \frac{q_0}{2\pi w}\frac{1}{r}, \tag{10}$$

where $r$ is the distance from the source point. Substitution (10) into (4) gives

$$R_\tau = k_\tau \left(\frac{t_n q_0}{2\pi w^2 r}\right)^n \frac{1}{|\partial w/\partial x|} \tag{11}$$

Equation (11) implies that the relative input of viscous shear tends to infinity, as well. Clearly, this is an artificial effect generated by the mathematical simplification of the injection process. Still, despite of the fact, that the average viscous shear over any symmetric vicinity of the singular point is zero, it may indicate that the particle velocities near the source may be great and, consequently, produce significant shear traction. Therefore, it might be of interest to estimate the distance, at which the velocity asymptotics (10) occurs. The accurate solutions of the axisymmetric problem for a penny-shaped fracture [8, 29] provide an option to make the evaluation.

For certainty, the fluid is assumed Newtonian ($n = 1$). Then the solution [8, 29] shows that the asymptotics (10) holds to the accuracy of 4.5 % at the distance of 0.155 $r_*$, where $r_*$ is a current fracture radius. For the radius $r_*$ exceeding 5 m, the distance $r = 0.155 r_* = 0.775$ m is notably greater than the standard borehole radius (some 0.1 m) what may justify using the approximation of the pointed source at such a distance.

A zone, having much less radius ($r = 0.01 r_*$), is certainly under the asymptotics (10). This size is notably less than a typical mesh-size $\Delta x$, used for modeling 3D fractures (see, e.g. [10, 12]). Therefore, if the ratio $R_\tau$ at the boundary of a zone of the size $0.01 r_*$ is small, then the shear traction, generated by the Dirac's delta-function, may be safely neglected in the elasticity equation. An estimation, which confirms that this is the case, is given in the next section, where the ratio $R_\tau$ is studied for the major part of a HF out of extremely close (less than usual mesh size) vicinities of the singular points. It is assumed that points at the distance $0.01 r_*$ from the *source (front)* are well within the asymptotic zone near the source (front).

**5. Input of viscous shear at the major part of HF**

The benchmark solutions to the plane-strain [7] and axisymmetric [8, 29] problems provide an opportunity to accurately calculate the ratio $R_\tau$ for the major part of the HF surface. In the both problems, the opening $w(x)$ and the particle velocity $v(x)$ entering (3) are defined via their self-similar counterparts $W(\varsigma)$ and $\gamma_x V(\varsigma)$ as (see, e.g. [29]):

$$w = \xi_* t_n^{n/(n+2)} t^{\gamma_w} W(\varsigma), \quad v = \xi_* \gamma_x t^{\gamma_x - 1} V(\varsigma), \quad \varsigma = x/x_*, \quad x_* = \xi_* t^{\gamma_x}, \quad \xi_* = \xi_{*n}[q_0/t_n^{n/(n+2)}]^{\gamma_q} \tag{12}$$

Herein, for the plane-strain problem, $\gamma_w = 1/(n+2)$, $\gamma_x = (n+1)/(n+2)$, $\gamma_q = 1/2$, $x_*$ is the fracture half-length; for axisymmetric problem, $\gamma_w = 1/3(2-n)/(n+2)$, $\gamma_x = 2/3(n+1)/(n+2)$, $\gamma_q = 1/3$, $x_*$ is the radius. Note that in the both cases, the difference $\gamma_x - \gamma_w = n/(n+2)$ is the same.

The constant $\xi_{*n}$ and the functions $W(\varsigma)$ and $V(\varsigma)$ ($0 \leq \varsigma \leq 1$) for various behaviour indices $n$ are known to the accuracy of five significant digits, at least. Then central differences provide accurate values of the derivative $dW/d\varsigma$ used below.

Employing (12) in (3) gives

$$R_\tau = k_\tau \left(\frac{t_n}{t}\right)^{\frac{n}{n+2}} f_n(\varsigma), \tag{13}$$

where

$$f_n(\varsigma) = \left(\frac{\gamma_x V(\varsigma)}{W(\varsigma)}\right)^n \frac{1}{|dW/d\varsigma|} \tag{14}$$

*Comment 2.* It is worth noting that the term on the right hand side of (14) is similar to the term on the right hand side of (3), when taking into account that $\gamma_x V(\varsigma)$, $W(\varsigma)$ and $dW/d\varsigma$ are self-similar counterparts of $v(x)$, $w(x)$ and $dw/dx$. The factor $\left(\frac{\gamma_x V(\varsigma)}{W(\varsigma)}\right)^n$ looks as a self-similar counterpart $2\mathcal{T}$ of the doubled shear traction $2\tau$ in (1). This leads to an illusion that the very ratio $f_n(\varsigma) = \frac{2\mathcal{T}}{|dW/d\varsigma|}$ itslef characterizes the relative input of hydraulically induced shear tractions as compared with the conventional term. This wrong suggestion, made in the paper [30] (Figures 1-3 of [30] and their discussion), drastically (four orders) overestimates the true input, defined by equation (13). The latter includes very small factor $(t_n)^{\frac{n}{n+2}}$, which, as shown below, makes $R_\tau$ much less than $f_n(\varsigma)$. This implies that the estimations of [30] are incorrect.

Evaluate the factor $(t_n/t)^{\frac{n}{n+2}}$ on the right hand side of (13). Clearly it is infinite at $t = 0$, what reflects instant application of non-zero pumping rate at the initial moment. For the time of one second ($t = 1$), this factor becomes quite small. Specifically, for a Newtonian fluid ($n = 1$) with typical dynamic viscosity $M = 0.1$ Pa s, and for typical rock modulus $E' = 25$ GPa, the factor is 3.63 $10^{-4}$. For a typical thinning fluid [31] ($n = 0.6$, $M = 0.39$ Pa s$^{0.6}$) and the same rock modulus, the factor is 1.44 $10^{-4}$. Importantly, it is of order $10^{-4}$.

The values of $f_n(\varsigma)$, entering (13), are much greater. Calculated by using the self-similar solution to the axisymmetric problem, they are given in Table for a Newtonian fluid ($n = 1$) and for a typical thinning fluid ($n = 0.6$).

| $\varsigma = r/r_*$ | 0.01 | 0.1 | 0.2 | 0.4 | 0.6 | 0.8 | 0.9 | 0.95 | 0.99 |
|---|---|---|---|---|---|---|---|---|---|
| $f_n(\varsigma)$ $n = 1$ | 21.5 | 1.33 | 0.544 | 0.111 | 0.144 | 0.121 | 0.130 | 0.152 | 0.246 |
| $f_n(\varsigma)$ $n = 0.6$ | 12.0 | 1.05 | 0.491 | 0.236 | 0.163 | 0.137 | 0.139 | 0.152 | 0.218 |

From the table it can be seen that in the major HF part ($0.1 \leq \varsigma \leq 0.95$), where the influence of singularities at the source ($\varsigma = 0$) and at the front ($\varsigma = 1$) is not extreme, the order of $f_n(\varsigma)$ does not exceed 1. Combining this estimation with that for the factor $(t_n/t)^{\frac{n}{n+2}}$, it is evident that at the major part of the fracture surface, for practically singnificant time ($t \geq 1$ s), the order of the ratio $R_\tau$ never exceeds $10^{-4}$. Moreover, even in a close vicinity ($\varsigma = 0.01$) of a pointed source, its order does not exceed $10^{-3}$. Similarly, on the boundary with the near front zone ($\varsigma \approx 0.95$) and even well inside this zone ($\varsigma = 0.99$), the order of the ratio $R_\tau$ is less than $10^{-4}$. The last estimation agrees with that made in Section 3 when considering the asymptotic equation (9) for the near-front zone. The agreement is due to the fact that the self-similar solution near the front meets the asymptotics (7) for the regime of dominating viscosity. It can be shown that under the asymptotic umbrella (7), eqn (13) reduces to the asymptotic eqn (9). The same decisive factor $t_n^{\frac{n}{n+2}}$ on the right hand sides of (9) and (13), which is of key significance for estimation of $R_\tau$, reflects this connection. Thus, as it should be, the estimation for the major part of the HF continuously transforms into the estimation for the near-front zone. The both yield the same order of $R_\tau$, not exceeding $10^{-4}$. Calculations for the plane-strain problem and for the limiting case of a perfectly plastic fluid, show that the estimations are true for them as well.

## 6. Summary

The analysis, presented in the paper, gives the quantitative estimation of the input of hydraulically induced shear traction into the elasticity equation for the entire surface of a HF fracture. It is established that, except for negligibly small vicinities of a pointed source and of the fluid front, the input is of order $10^{-4}$ of that of the conventionally accounted term. Near the front, it reaches the level of 1% merely at the distance of atomic size. Near a pointed source, it is still of order 0.1% at a distance less than the borehole size. This distance, being less than a reasonable mesh size used in numerical simulations not specially designed to model the flow through perforation, the input is negligible near the source, as well.

Therefore, the conclusion of the papers [14, 15], that the influence of the shear traction may be confidently neglected in the near-front zone, is extended to the entire fracture surface. This implies that there is no need to account for them not only in the *form* of a fracture condition, but also in *factors* entering conventional near-front asymptotics and fracture conditions.

**Acknowledgement.**
The author appreciates the support of the Polish National Scientific Center (Grant No. 2015/19B/ST8/00712).

**Appendix. 3D elasticity equation accounting for hydraulically induced shear traction**

The boundary integral equations for piece-wise homogeneous isotropic elastic medium, composed of blocks with the same Poisson's ratio $\nu$ and possibly different shear modules $\mu$, are given in the paper [3] (see also [4, 5]). The blocks may contain inclusions, cracks and cavities. For our purpose, we employ the hypersingular equation, which defines tractions via the displacement discontinuities:

$$-\int_S \frac{1}{2\mu} J_H(x,\xi) \Delta u(\xi) dS_\xi + \int_S \left( \frac{1}{2\mu^+} J_S^+(x,\xi) t_n^+(\xi) - \frac{1}{2\mu^-} J_S^-(x,\xi) t_n^-(\xi) \right) dS_\xi = \frac{1}{2} \left( \frac{1}{2\mu^+} t_n^+(x) + \frac{1}{2\mu^-} t_n^-(x) \right) \quad x \in S \quad (A1)$$

Herein, $S$ is the entire surface, at which a discontinuity of the displacement vector $\Delta u = u^+ - u^-$ and/or traction vector $\Delta t_n = t_n^+ - t_n^-$ occurs. The normal $n$ is fixed arbitrary on a contact of adjacent blocks, on cracks and inclusions. The index "plus" ("minus") refers to the limiting value from the side with respect to which the normal $n$ is outward (inward). For the external boundaries, the normal is assumed to be outward; then $u^- = 0$ and $t_n^- = 0$. The singular matrix $J_S(x,\xi)$ is obtained by applying the traction operator $T_{nx}$ to the columns of the matrix $U(x,\xi)$ of fundamental solutions: $J_S(x,\xi) = T_{nx} U(x,\xi)$. The hypersingular matrix $J_H(x,\xi)$ is defined as $J_H(x,\xi) = T_{nx} \left( T_{n\xi} U(\xi,x) \right)^T$.

When taking $U(x,\xi)$ as that corresponding to the Kelvin's solutions, the entries of $U(x,\xi)$, $J_S(x,\xi)$ and $J_H(x,\xi)$ are: $U_{ij}(x,\xi) = \frac{1}{16\pi\mu(1-\nu)} \left[ 4(1-\nu) \frac{1}{R} \delta_{ij} - \frac{\partial^2 R}{\partial x_i \partial x_j} \right]$,

$$J_{Sij}(x,\xi) = \frac{1}{8\pi(1-\nu)}\left\{\left[2\nu\delta_{ik}\frac{\partial}{\partial x_j} + 2(1-\nu)\left(\delta_{ij}\frac{\partial}{\partial x_k} + \delta_{jk}\frac{\partial}{\partial x_i}\right)\right]\frac{1}{R} - \frac{\partial^3 R}{\partial x_i \partial x_j \partial x_k}\right\}n_{xk} \quad (A2)$$

$$J_{Hij}(x,\xi) = \frac{\mu}{4\pi(1-\nu)}\left\{n_{\xi m}\frac{\partial^4 R}{\partial x_m \partial x_i \partial x_j \partial x_k} - \left[2\nu\left(\delta_{ik}n_{\xi m}\frac{\partial^2}{\partial x_m \partial x_j} + n_{\xi j}\frac{\partial^2}{\partial x_i \partial x_k}\right) + (1-\nu)\left(n_{\xi i}\frac{\partial^2}{\partial x_j \partial x_k} + n_{\xi k}\frac{\partial^2}{\partial x_i \partial x_j} + \delta_{ij}n_{\xi m}\frac{\partial^2}{\partial x_m \partial x_k} + \delta_{kj}n_{\xi m}\frac{\partial^2}{\partial x_m \partial x_i}\right)\right]\frac{1}{R}\right\}n_{xk}, \quad (A3)$$

where $R = \sqrt{(x_i - \xi_i)^2}$ is the distance between an integration $\xi$ and field $x$ points; $n_\xi$ and $n_x$ are the unit normals to the surface $S$ at the point $\xi$ and $x$, respectively; $\delta_{ik}$ is the Kronecker delta. Summation over a repeated Latin index is assumed in (A2) and (A3).

Consider the case of a *planar* hydraulic fracture with the normal $n$. Direct the axis $x_1$ of the global system in the direction of $n$. Then $n_{xk} = \delta_{1k}$, $n_{\xi m} = \delta_{1m}$, and for the normal component ($i = 1$) of the traction on the right hand side of (A1), equations (A2) and (A3) give:

$$J_{S1j}(x,\xi) = \frac{1}{8\pi(1-\nu)}(1-2\nu)\frac{x_j - \xi_j}{R^3}, \quad J_{Hij}(x,\xi) = \frac{\mu}{4\pi(1-\nu)}\delta_{1j}\frac{1}{R^3} \quad (A4)$$

For a homogeneous medium ($\mu^+ = \mu^+ = \mu$) with a planar surface $S$ being the surface of a hydraulic fracture, the terms on the right hand side of (A1), corresponding to the normal component, are $t_{n1}^+(x) = t_{n1}^-(x) = -p$, where $p$ is the net-pressure of a fracturing fluid. Then using (A4) in (A1) leads to the general elasticity equation, defining the net-pressure via the fracture opening $w = -\Delta u_1 = u_1^- - u_1^+$ and components $\Delta t_{n2} = t_{n2}^+ - t_{n2}^-$, $\Delta t_3 = t_{n3}^+ - t_{n3}^-$ of the discontinuity of the hydraulically induced shear traction. Since for these tractions, $t_{n2}^+ = -t_{n2}^-$, $t_{n3}^+ = -t_{n3}^-$, the discontinuities are $\Delta t_{n2} = 2\tau_2$, $\Delta t_{n3} = 2\tau_3$, where $\tau_2$ and $\tau_3$ are components of the hydraulically induced shear traction on that fracture side, with respect to which the normal is outward. Then using (A4) in (A1) yields

$$\frac{E'}{8\pi}\int_S \frac{w(\xi)}{R^3}dS_\xi + \frac{1-2\nu}{8\pi(1-\nu)}\int_S \left(2\tau_2\frac{x_2 - \xi_2}{R^3} - 2\tau_3\frac{x_3 - \xi_3}{R^3}\right)dS_\xi = -p \quad x \in S, \quad (A5)$$

where $E' = 2\mu/(1-\nu)$ is the plane-strain elasticity modulus (equivalently, $E' = E/(1-\nu^2)$).

The first integral on the right hand side of (A5) is always accounted for in papers on HF (e.g. 8-10, 12, 29]). The second integral, corresponding to the hydraulically induced shear traction, is neglected.

To compare their relative input into the elasticity equation, we transform the first integral to the form of the second. The identity $\frac{\partial^2}{\partial x_1^2}\frac{1}{R} = -\frac{1}{R^3} + 3\frac{x_1 - \xi_1}{R^5}$ implies that on the fracture plane ($x_1 = \xi_1 = 0$) $\frac{1}{R^3} = -\frac{\partial^2}{\partial x_1^2}\frac{1}{R}$. Since the function $\frac{1}{R}$ is harmonic $\left(\frac{\partial^2}{\partial x_1^2}\frac{1}{R} + \frac{\partial^2}{\partial x_2^2}\frac{1}{R} + \frac{\partial^2}{\partial x_3^2}\frac{1}{R} = 0\right)$, this yields $\frac{1}{R^3} = \frac{\partial^2}{\partial x_2^2}\frac{1}{R} + \frac{\partial^2}{\partial x_3^2}\frac{1}{R}$. In view of the definition of $R$, $\frac{\partial R}{\partial x_k} = -\frac{\partial R}{\partial \xi_k}$. Then on the fracture plane, $\frac{1}{R^3} = -\frac{\partial^2}{\partial x_2 \partial \xi_2}\frac{1}{R} - \frac{\partial^2}{\partial x_3 \partial \xi_3}\frac{1}{R}$. After substitution this equation into (A5) and using the Gauss theorem, we obtain

$$\frac{E'}{8\pi}\left(\int_S \left[\left(\frac{\partial w(\xi)}{\partial \xi_2} - \frac{1-2\nu}{E'(1-\nu)}2\tau_2(\xi)\right)\frac{x_2 - \xi_2}{R^3} + \left(\frac{\partial w(\xi)}{\partial \xi_3} - \frac{1-2\nu}{E'(1-\nu)}2\tau_3(\xi)\right)\frac{x_3 - \xi_3}{R^3}\right]dS_\xi = p \quad x \in S \quad (A6)$$

At each point of the fracture surface $S$, the relative input of the shear traction may be compared in the system with the $\xi_2$-axis in the direction $\xi_\tau$ of the fluid velocity. Then in (A6) $\tau_2$ equals the magnitude $\tau$ of the shear traction, while $\tau_3 = 0$. Hence, for a planar fracture, the relative input of the shear traction is defined by the ratio

$$R_\tau = \frac{2k_\tau \tau}{E'|\partial w/\partial \xi_\tau|}, \quad (A7)$$

where $k_\tau = \frac{1-2\nu}{1-\nu}$. Equation (A6) is equation (2), in which the symbol $x$ replaces $\xi_\tau$.

**References**

1. Spence D.A., Sharp P.W. Self-similar solutions for elastohydrodynamic cavity flow // *Proc. Roy Soc. London. Ser. A*. 1985. V. 400. N 819. P. 289-313.
2. Muskhelishvili N.I. *Some Basic Problems of the Mathematical Theory of Elasticity*. Groningen, Noordhoff. 1975. 746 p.
3. Linkov A. Real and complex hypersingular integrals and integral equations in computational mechanics // *Demonstratio Mathematica*. Politechnika Warszawska. 1995. V. 28. No 4. P. 759-769.
4. Linkov A. *Boundary Integral Equations in Elasticity Theory*. Kluwer Academic Publishers, Dordrecht-Boston-London. 2002. 268 p.
5. Jaworski D., Linkov A.M., Rybarska-Rusinek L. On solving 3D elasticity problems for inhomogeneous region with cracks, pores and inclusions // *Computers and Geotechnics*. 2016. V. 71. P. 295-309.
6. Lenoach B. The crack tip solution for hydraulic fracturing in a permeable solid // *J. Mech. Phys. Solids*. (1995). V. 43. P. 1025-1043.
7. Adachi J.I., Detournay E. Self-similar solution of plane-strain fracture driven by a power-law fluid // *Int. J. Numer. Anal. Meth. Geomech*. 2002. V. 26. P. 579-604.
8. Savitski A., Detournay E. Propagation of a fluid driven penny-shaped fracture in an impermeable rock: asymptotic solutions // *Int. J. Solids Struct*. 2002. V. 39. P. 6311-6337.
9. Adachi J., Siebrits E., Pierce A., Desroches J. Computer simulation of hydraulic fractures // *Int. J. Rock Mech. Mining Sci*. 2007. V. 44. P. 739-757.
10. Peirce A., Detournay E. An implicit level set method for modeling hydraulically driven fractures // *Comput. Methods Appl. Mech. Engng*. 2008. V. 197. P. 2858-2885.
11. Linkov A.M. Particle velocity, speed equation and universal asymptotics for efficient modelling of hydraulic fractures // *J. Appl. Math. Mech*. 2015. V. 7. P. 54-63.
12. Peirce A. Implicit level set algorithms for modeling hydraulic fracture propagation // Phil. Trans. Roy. Soc. A. 2016. 374: 20150423.
13. Wrobel M., Mishuris G., Piccolroaz A. Energy release rate in hydraulic fracture: Can we neglect an impact of the hydraulically induced shear stress? // *Int. J. Eng.Sci*. 2017. V. 111. P. 28-51.
14. Linkov A.M. On influence of shear traction on hydraulic fracture propagation // *Materials Physics and Mechanics*. 2017. V. 32. P. 272-277.
15. Linkov A.M. Response to the paper by M. Wrobel, G. Mishuris, A. Piccolroaz "Energy release rate in hydraulic fracture: Can we neglect an impact of the hydraulically induced shear stress?" (International Journal of Engineering Science, 2017, 111, 28–51) // *Int. J. Eng.Sci*. 2018. V. 127, 217-219.
16. Kemp L.F. Study of Nordgren's equation of hydraulic fracturing // *SPE Production Eng*. 1990. V. 5 (SPE 19959). P. 311-314.
17. Linkov A.M. 2011. Use of a speed equation for numerical simulation of hydraulic fractures // *Available online at: http://arxiv.org/abs/1108.6146*. 2011. Date: Wed, 31 Aug 2011 07:47:52 GMT (726kb). Cite as: arXiv: 1108.6146v1 [physics.flu-dyn].
18. Linkov A.M. On efficient simulation of hydraulic fracturing in terms of particle velocity // *Int. J. Eng. Sci*., 2012. V. 52. P. 77-88.
19. Linkov A.M. Universal asymptotic umbrella for hydraulic fracture modeling // *Available online at http://arxiv.org/abs/1404.4165.* 2014. Date: Wed, 16 Apr 2014 08:35:16 GMT (564kb) Cite as: arXiv: 1404.4165 [physics.flu-dyn]. 11 p.
20. Rice J.R. Mathematical analysis in the mechanics of fracture // H. Liebowitz (Ed.), *Fracture, an Advanced Treatise*. V. II. New York, Academic Press. 1968. P. 191-311.
21. Descroches J., Detournay E., Lenoach B. et al. The crack tip region in hydraulic fracturing // *Proc. Roy. London. Ser. A*. 1994. V. 447. N. 1929. P. 39-48.

22. Garagash D.I. Transient solution for a plane-strain fracture driven by a shear-thinning, power-law fluid // *Int. J. Numerical and Analytical Methods in Geomechanics*. 2006. V. 30. P. 1439-1475.

23. Garagash D.I., Detournay E., Adachi J.I. Multiscale tip asymptotics in hydraulic fracture with leak-off // *J. Fluid Mech*. 2011. V. 669. P. 260-297.

24. Gordeliy E., Peirce A. Implicit level set schemes for modeling hydraulic fractures using the XFEM // *Computer Meth. Appl. Mech. Eng*. 2013. V. 253. P. 125-143.

25. Neuber H. *Theory of Notch Stresses*. Ann Arbor, Michigan. J.W. Edwards. 1946.

26. Novozhilov V.V. On a necessary and sufficient criterion for brittle fracture // *J. Appl. Math. Mech*. 1969. V. 33. No 2. P. 201-210.

27. Dobroskok A.A., Ghassemi A., Linkov A.M. Extended structural criterion for numerical simulation of crack propagation under compressive loads // *Int. J. Fracture*. 2005. V. 133. P. 223-246.

28. Linkov A.M. Loss of stability, characteristic length and Novozhilov-Neuber criterion in fracture mechanics // *Mechanics of Solids*. 2010. No 6. P. 844-854.

29. Linkov A.M. Solution of axisymmetric hydraulic fracture problem for thinning fluids // *J. Appl. Math. Mech*. 2016, V. 8. No 2. P. 149-155.

30. Wrobel M., Mishuris G., Piccolroaz A. (2017). On the impact of tangential traction on the crack surfaces induced by fluid in hydraulic fracture: response to the letter of A.M. Linkov. Int. J. Eng. Sci. (2018) 127, 217-219 // *Int. J. Eng.Sci.* 2018. V. 127, 220-224.

31. Montgomery C. Key-Note lecture: Fracturing fluids // *Effective and Sustainable Hydraulic Fracturing*. A. P. Bunger, J. McLennan, R. Jeffrey (eds). 2013. Published by InTech (Croatia) P. 3-24. Proc. Int. Conference HF-2013. Available on line: www.intechopen.com.